# Tailoring Magnetism in Self-intercalated $Cr_{1+\delta}Te_2$ Epitaxial Films

Y. Fujisawa[1], M. Pardo-Almanza[1], J. Garland[1,2], K. Yamagami[1], X. Zhu[1], X. Chen[3], K. Araki[4],
T. Takeda[4], M. Kobayashi[4,5], Y. Takeda[6], C. H. Hsu[7], F. C. Chuang[7], R. Laskowski[8],
K. H. Khoo[8], A. Soumyanarayanan[3,9], Y. Okada[1]

[1]*Quantum Materials Science Unit, Okinawa Institute of Science and Technology (OIST),
Okinawa 904-0495, Japan*
[2] *Applied Physics & Mathematics Department, Northeastern University,
Boston, Massachusetts 02115, USA*
[3] *Institute of Materials Research and Engineering, Agency for Science Technology and Research, 138634 Singapore*
[4] *Department of Electrical Engineering and Information Systems, The University of Tokyo, 7-3-1 Hongo,
Bunkyo-ku, Tokyo 113-8656, Japan*
[5] *Center for Spintronics Research Network, The University of Tokyo, 7-3-1 Hongo, Bunkyo-ku, Tokyo 113-8656, Japan*
[6] *Condensed Matter Science Division, Quantum Beam Science Directorate,
Japan Atomic Energy Agency, 1-1-1 Kouto, Sayo-cho, Sayo-gun, Hyogo 679-5148, Japan*
[7] *Department of Physics, National Sun Yat-sen University, Kaohsiung 80424, Taiwan*
[8] *Institute of High Performance Computing, Agency for Science, Technology and Research,
138632 Singapore*
[9]*Department of Physics, National University of Singapore, 117551 Singapore*

Magnetic transition metal dichalcogenide (TMD) films have recently emerged as promising candidates to host novel magnetic phases relevant to next-generation spintronic devices. However, systematic control of the magnetization orientation, or anisotropy, and its thermal stability, characterized by Curie temperature ($T_c$) – remains to be achieved in such films. Here we present self-intercalated epitaxial $Cr_{1+\delta}Te_2$ films as a platform for achieving systematic/smooth magnetic tailoring in TMD films. Using a molecular beam epitaxy (MBE) based technique, we have realized epitaxial $Cr_{1+\delta}Te_2$ films with smoothly tunable δ over a wide range (0.33-0.82), while maintaining NiAs-type crystal structure. With increasing δ, we found monotonic enhancement of $T_c$ from 160 to 350 K, and the rotation of magnetic anisotropy from out-of-plane to in-plane easy axis configuration for fixed film thickness. Contributions from conventional dipolar and orbital moment terms are insufficient to explain the observed evolution of magnetic behavior with δ. Instead, *ab initio* calculations suggest that the emergence of antiferromagnetic interactions with δ, and its interplay with conventional ferromagnetism, may play a key role in the observed trends. To our knowledge, this constitutes the first demonstration of tunable $T_c$ and magnetic anisotropy across room temperature in TMD films, and paves the way for engineering novel magnetic phases for spintronic applications.

**INTRODUCTION**

Epitaxial thin films and heterostructures have provided pristine platforms for exploring novel functionalities and phenomena [1,2]. Meanwhile, the concomitant development of material "knobs" to tailor these properties is key to exploit them for technological applications [3,4]. For example, the development of magnetic thin films for high-performance memory applications has required systematic modulation of the magnetic anisotropy – which sets the device geometry and switching characteristics - and enhancement of the temperature scale governing magnetic stability [5,6]. Recently, magnetic anisotropy has also served as an important knob in realizing exotic topological states in thin films [7,8,9,10,11], wherein practical considerations also require efforts to achieve magnetic stability beyond room temperature [9]. These efforts motivates the development of epitaxial thin film platforms wherein the magnetic properties may be smoothly modulated, and also serve the search for novel magnetic states.

Recent intensive studies point to transition metal chalcogenides (TMDs) as one of the most promising platforms to realize such tunable characteristics [12,13]. $Cr_{1+\delta}Te_2$ in particular is a unique self-intercalated TMD – expected to have rich magnetic properties based on bulk investigations [14,15,16,17,18,19,20,21,22,23,24,25,26]. Previous studies of bulk crystals grown with varying compositions e.g. CrTe [14,15,16], $Cr_3Te_4$ [17], $Cr_2Te_3$ [18,26], $Cr_5Te_8$ [19,20,21,22], and $CrTe_2$ [23,24,25] - were shown to have the same NiAs-type structure, with ferromagnetic transition temperature ($T_C$) ranging between 170 to 350 K. The notation $Cr_{1+\delta}Te_2$ conveniently represents the compositions of all these compounds, wherein δ is the fraction of Cr atoms self-intercalated between neighboring $CrTe_2$ layers (**see Figure 1a**). Notably, in addition to conventional ferromagnetic (FM) interaction, $Cr_{1+\delta}Te_2$ bulk crystals are suggested to host antiferromagnetic (AF) interactions, non-collinear spin textures, and tunable magnetic anisotropy [27,28,29,30]. Meanwhile, epitaxial thin film growth has focused on $Cr_2Te_3$(001), due to its perpendicular magnetic anisotropy with $T_c$ ~ 170 K [31,32,33,34,35], and recent studies have suggested the presence of emergent topological phenomena in ultra-thin $Cr_2Te_3$ films or interfaces [34,36,37,38]. However, the absence of means to systematically control δ has constrained a deeper understanding of underlying magnetic interactions in $Cr_{1+\delta}Te_2$, and limited the potential for engineering magnetic phases. Notably, the intrinsically high vapor pressure of Te (and other chalcogens in general), has limited efforts to vary δ over a wide enough range using conventional epitaxial film growth methods.

Here, we establish a method to realize epitaxial $Cr_{1+\delta}Te_2$ films with smoothly modulated δ over a wide range (0.33-0.82), while maintaining the same crystal structure, with using molecular beam epitaxy (MBE) based growth technique. We successfully tuned the Curie temperature $T_c$ from 160 K to beyond 350 K, and the magnetic anisotropy smoothly between out-of-plane (OP) and in-plane (IP) easy axes. Our calculations suggest that the modulation of magnetic properties originates from the introduction of AF interactions due to self-intercalated Cr atoms, and their interplay with conventional FM. This establishes $Cr_{1+\delta}Te_2$ as a promising epitaxial platform to investigate unconventional magnetic phases.

**EXPERIMENTAL RESULTS**

The growth of epitaxial $Cr_{1+\delta}Te_2$ films with different δ on $Al_2O_3$ (001) substrate was achieved by following a two step-procedure using molecular beam epitaxy (MBE): (i) preparation of as-grown $Cr_{1+\delta}Te_2$ films with fixed δ, and (ii) additional in-situ post-deposition annealing (iPDA) to tune δ. First, we describe the procedure used to prepare as-grown epitaxial films. We began by preparing a clean, atomically flat $Al_2O_3$ (001) substrate surface whose quality is verified using reflection high energy electron diffraction (RHEED) (**Figure 1b**). $Cr_{1+\delta}Te_2$ films were then deposited at a substrate temperature of 300 ℃, with as-grown film thickness consistently kept to ~80 nm in this work. An atomically flat surface with negligible roughness on the as-grown film was confirmed by sharp RHEED streaks (**Figure 1c**) and in-situ scanning tunneling microscope (STM) topographic image (**Figure 1d**). Notably, the stoichiometric composition of the as-grown film was found to be δ = 0.33 using ex-situ energy dispersive X-ray spectroscopy (EDS) characterization. The ex-situ structural characterization using X-ray diffraction (XRD), shown for the (00*L*) direction, does not indicate the presence of any impurity phases (**Figure 1e**). Furthermore, based on an azimuthal XRD scan (**Figure 1f**), the epitaxial nature of the films was confirmed by establishing the crystallographic orientation of the film relative to the substrate (**Figure 1g**).

iPDA enables systematic control of the high volatility element – Te in our case- while keeping film surface atomically flat [39,40,41,42]. The iPDA process involves annealing the as-grown film under adequate flow of Te flux for 30 minutes at a substrate temperature ($T_A$). In this work, $T_A$ was varied between 300 and 700 ℃ for different iPDA sequences while keeping all other growth parameters unchanged (**see SI for details**). **Figure 2a** shows the ex-situ XRD profiles for films with iPDA at varying $T_A$ values. For samples with $T_A$ ranging over 400 – 600 C, all peaks can be indexed by either substrate or $Cr_{1+\delta}Te_2$ (00*L*). Meanwhile, samples with $T_A$ = 650 and 700 ℃ show additional peaks identified as originating from a pure Cr phase (arrows in **Figure 2a**). While the $Cr_{1+\delta}Te_2$ (003) peak is proximate to the additional Cr(210) peak, systematic investigation of EDS and XRD data enables a clear disentanglement of these two peaks (see SI for more details). Thus, we conclude that $T_A$ ~ 650 ℃ corresponds to the temperature above which the film may degrade into multiple crystalline phase. Importantly, for $T_A$ below ~650 ℃, the relation between $T_A$ and δ, determined from EDS, indicates a monotonic and smooth increase (**Figure 2b**). Based on the empirical δ($T_A$) relation, we conclude that near-complete CrTe (δ~1) can be obtained between 525 ℃ and 600 ℃. Here the decrease of (003) peak intensity can be explained by the extinction rule for ideal CrTe (δ = 1) with NiAs-type crystal structure – where (003) peak cannot exist[43]. Meanwhile, the (003) peak may be prominent for smaller δ when the crystal structure has two inequivalent layers (see SI for details). In summary, our iPDA technique realizes systematic tuning of δ while maintaining the crystal structure of the as-grown $Cr_{1+\delta}Te_2$ film.

Structural characterization further indicates the monotonic change of lattice parameters with varying δ. **Figure 2c** shows the δ-dependence of lattice constant along IP ($a_0$) and OP ($c_0$), as well as the unit volume $V_0$ ( $= \frac{\sqrt{3}}{2} a_0^2 c_0$). Importantly, based on Vegard's law [44], the linear relation between lattice parameters and δ confirms the systematic change in the self-intercalation of Cr atoms between neighboring $CrTe_2$ layers (see **Figure 1a**). Finally, we compare the values of $a_0$ and $c_0$ for our films with literature on bulk crystals with similar δ (**Figure 2c** upper panel). We find that our films have a slightly elongated $a_0$ and correspondingly shorter $c_0$ compared to bulk crystals [14,16,19] – resulting in a similar unit cell volume (**Figure 2c**). This points to the existence of a finite IP tensile strain in our films, which could modify the electronic and magnetic properties with respect to bulk crystals.

The controlled stoichiometric variation within $Cr_{1+δ}Te_2$ films should produce measurable effects on the magnetic properties – investigated here using magnetization (*M*) measurements - across temperatures M(T) (**Figure 3a**) and external fields M(H) (**Figure 3b**). First, we present the δ-dependence of the *T*c by examining field-cooled (*H* = 1000 Oe) M(T) curves along IP (red) and OP (blue) orientations. As seen clearly in **Figure 3a** across both IP and OP M(T) curves, $T_C$ increases monotonically with δ. The systematic evolution of $T_C$ with δ is summarized in **Figure 3c (left axis)**: smoothly increasing from ~160 K for δ = 0.33 **(top in Figure 3a)** up to ~350 K for δ = 0.82 **(bottom in Figure 3a)**. Notwithstanding prolific research activities on epitaxial $Cr_{1+δ}Te_2$ films, this is – to our knowledge – the first demonstration of systematically tunable $T_C$ across room temperature.

More interestingly, we also see a systematic evolution of magnetic anisotropy with δ. The anisotropy change is evident from the change in remnant magnetization of the IP and OP M(T) curves around δ = 0.4 samples (**Figure 3a**). For a quantitative picture, we turn to M(H) hysteresis curves acquired across δ at T = 2 K in IP and OP configurations (**Figure 3b**). For each δ, the effective anisotropy ($K_{eff}$) is determined by the areal difference between the IP and OP M(H) curves [45]. The variation of $K_{eff}$ with δ is summarized in **Figure 3d (left axis)** and is consistent with the expected trend from *M*(*T*) curves in **Figure 3a**. For δ=0.4 and below, $K_{eff}$ is positive which signifies OP, or perpendicular magnetic anisotropy. However, $K_{eff}$ reduces with increasing δ and changes sign across δ~0.4, resulting in IP magnetic anisotropy for δ>0.4.

Typically sample shape and magnetocrystalline (or uniaxial) contributions play major roles in determining magnetic anisotropy [45]. Here we examine the expected contributions from these two effects. First, shape anisotropy originates from magnetostatic, or dipole interactions, which for thin films leads to preferential IP anisotropy. However, such effects are consistent across δ, and therefore cannot describe the observed evolution with δ (**see SI**). Meanwhile, magnetocrystalline contributions arise from the atomic orbital moments. To examine this effect, we performed X-ray magnetic circular dichroism (XMCD) measurements at Cr $L_{2,3}$ edge, and find that the orbital moment is negligibly small and does not vary measurably with δ (**see SI**). Thus, neither of the conventional anisotropy contributions – shape or uniaxial – can explain the observed anisotropy switching with δ.

**THEORETICAL CALCULATIONS**

To elucidate the physics governing the observed doping evolution of magnetism in $Cr_{1+\delta}Te_2$ films, we performed DFT calculations [39,40,41,42] for the stoichiometric compounds with δ = 0 and δ = 1, i.e. $CrTe_2$ (**Figure 4a and c**) and CrTe (**Figure 4b and d**). The exchange interactions for these compounds were calculated using the frozen magnon method [43]. Within this technique, real space exchange couplings are determined from the Fourier transform of the k-space dispersion of spin spirals (**methods and SI for details**). Importantly, while δ = 0 ($CrTe_2$) has only one Cr atom per unit cell, δ = 1 (CrTe) has two Cr sublattices within its basis (**Figure 4a-b**). Thus, for δ = 0, we have only one relevant interaction – the intra-sublattice exchange interaction, $J_{11}$ (green arrows in **Figure 4a**). In contrast, δ = 1 has two distinct intra-sublattice exchange interactions $J_{11}$ and $J_{22}$ and additionally, an *inter*-sublattice exchange interaction $J_{12}$ (red arrows in **Figure 4b**). From inspecting **Figure 4b**, we conclude that the two intra-sublattices for δ = 1 are physically indistinguishable, and therefore $J_{11}$ and $J_{22}$ should behave similarly. As a consistency check, we find that our DFT calculations correctly reproduce this similarity (**Figure 4d**), and are also in line with previous calculations for individual dopings [21,22,23,24].

**Figure 4c-d** compare the spatial decay of the 3 intra- and one inter-sublattice exchange interaction for δ = 0,1 respectively. First, a comparison of the intra-sublattice interactions shows that $J_{11}$ and $J_{22}$ (δ = 1) are identical, and show very similar spatial decay to $J_{11}$ (δ = 0). This may be associated with a ~+5meV nearest neighbor FM interaction – expected to be consistent across δ (**Figure 4c and d**). Meanwhile, the inter-sublattice exchange $J_{12}$ arises in δ = 1 from the addition of a full doping layer and contrasts strongly with the radial profiles of $J_{11}$ and $J_{22}$. Crucially, $J_{12}$ shows a mix of FM and AF interactions – evolving from strongly negative (-10meV) for the nearest neighbor to moderately positive (+3 meV) for next nearest neighbor, highlighted by shaded pink line in **Figure 4d**. Taken together, the DFT results for δ = 0,1 suggest a competition between the AF nearest neighbor inter-sublattice $J_{12}$=-10meV and FM intra-sublattice $J_{11}$=$J_{22}$=+5meV and next nearest neighbor inter-sublattice $J_{12}$=+3meV. The interplay of these interactions is vital to explaining for the observed evolution of magnetic properties.

Next, we proceed to compare the experimental and DFT results to elucidate the doping evolution of magnetism in $Cr_{1+\delta}Te_2$. First, the calculated exchange interactions, under the random phase approximation (RPA), give Curie temperatures, $T_C$=357K for δ=0 ($CrTe_2$) and $T_C$=491K for δ=1 (CrTe). These results are qualitatively consistent with the experimental trend. In this light, the monotonic enhancement of Tc with doping (**Figure 3b)** can be understood to result from the increased effective exchange field strength at Cr sites due to additional inter-sublattice interactions $J_{12}$ with intercalated Cr atoms[46]. The effective exchange field is determined by summing exchange interactions from all neighboring sites. Furthermore, the experimentally observed anisotropy evolution with δ may also be interpreted in view of the calculated exchange interactions. **Figure 4e** shows a possible schematic for the doping evolution of alignment of local moments in $Cr_{1+\delta}Te_2$. While moments in the original layers are aligned FM (light-blue arrows in **Figure 4e**), those in the self-intercalated layer (pink arrows in **Figure 4e**) tend to align AF with respect to the original layers – which could result in non-collinearity. Indeed, such non-collinearity is consistent with our calculations (Fig. S7), and also seen in previous literature, especially in high δ compounds [28, 29]. The non-collinear state in Fig. 4e provides a viable explanation of the observed doping evolution of anisotropy. This picture also supports the observation of coercive hysteresis for both IP and OP orientations across the entire doping range 0.33<δ<0.82, which suggests the existence of net moments within both orientations (**Figure 3c**).

The tunable magnetism realized within our $Cr_{1+\delta}Te_2$ films provides a promising platform for burgeoning efforts in topological magnetism [36,37,38]. Modulating δ alters not only the spacing between self-intercalated Cr atoms, but also the Fermi surface volume and geometry. This provides a rich playground for magnetic interactions, including direct FM and AF exchange, super-exchange, and Ruderman-Kasuya-Kittel-Yosida (RKKY) interactions leading to qualitatively different intra- and inter-sublattice exchange couplings. Interestingly, competition between direct FM exchange and higher order AF exchange interactions in Pd/Fe/Ir(111)[47], and RKKY-like interactions in $Gd_2PdSi_3$[48] is expected to stabilize magnetic skyrmions. Recently much excitement has centered around using the interfacial Dzyaloshinskii-Moriya interaction (i-DMI) to generate topological spin textures in magnetic thin films. Consequently, magnetic TMD films are being engineered to similarly achieve i-DMI [8,9,36,37,38]. Contrastingly, our work shows that epitaxial $Cr_{1+\delta}Te_2$ films may host other novel competing interactions – and could host unexplored emergent ground states.

**SUMMARY**

In summary, we have established a growth technique to controllably vary the fraction of self-intercalated Cr atoms within epitaxial $Cr_{1+\delta}Te_2$ films while maintaining the original crystal structure. As a consequence, we have realized tunable magnetism – including $T_c$ beyond room temperature and smooth modulation of magnetic anisotropy between OP and IP configurations. These effects are expected to arise from the interplay of FM and AF interactions between Cr atoms and bode well for the imminent applicability of epitaxial $Cr_{1+\delta}Te_2$ films.

**ACKNOWLEDGEMENTS**

The crystals structures of $Cr_{1+\delta}Te_2$ was visualized using VESTA [49]. This work is partly supported by Japan Society and Science and Technology Agency (JST) Core Research for Evolution Science and Technology (CREST), Japan, Grant number JPMJCR1812. This work was also partially supported by the Spintronics Research Network of Japan (Spin-RNJ). Supporting experiments at SPring-8 were approved by the Japan Synchrotron Radiation Research Institute (No. 2019B3841), nanoplatform (No. A-19-AE-0040), and Japan Atomic Energy Agency (No. 2019B-E20). The work in Singapore was supported by the SpOT-LITE program (Grant No. A18A6b0057), funded by Singapore's RIE2020 initiatives. We acknowledge the support of the National Supercomputing Centre (NSCC), Singapore for computational resources.

**METHODS**

The films were grown using molecular beam epitaxy (MBE) equipment, whose base pressure was ~$10^{-10}$ Torr. Cr and Te were evaporated using an e-beam evaporator and a K-cell. An electron beam gun and screen are equipped to check the surface quality of the sample. The $Cr_{1+\delta}Te_2$ films were deposited on $Al_2O_3$ substrates with area of approximately 3 x 4 $mm^2$. Regarding the detail growth sequence, see the followings and the supplemental information. In-situ STM observation has been carried out at 10 K with using ultra-high vacuum STM. Films were transferred from the MBE chamber to the STM chamber without exposing films to the air, using an ultra-high vacuum suitcase with its base pressure of $10^{-10}$ torr. In order to get electric contact between the film and bias electrodes for STM observation, Ti and Pt were deposited on the edge of $Al_2O_3$ substrate before MBE thin film growth. The X-ray diffraction measurements were performed at room temperature and ambient pressure using Bruker D8 discover. Cu K$\alpha$1 ($\lambda$ = 1.5418 nm) is used as the X-ray source. The measurements were done at room temperature. Lattice constant $c_0$ is estimated from XRD profiles along (00L) direction, while (101) diffraction is used to estimate lattice constant $a_0$. After calculating $d_{101}$ from these data, $a_0$ is calculated from $d_{101}$ and $c_0$ based on the hexagonal symmetry of crystals (see **Fig. S2**). The chemical composition was determined based on EDS measurement using FEI Quanta 250 FEG. To protect against surface degradation, before all of ex-situ measurements shown in this report, films were capped with amorphous Se layer at room temperature with 50~100nm thickness after cooing sample more than 10 hours. DC magnetization measurements were performed using Magnetic Property Measurement System (MPMS®3, Quantum Design). First-principles DFT calculations were performed using the WIENNCM package that employs linearized augmented plane waves (LAPW) as a basis set [50,51,52,53]. $R_{kmax}$ was set to 7.5 for the energy cutoff and, the Brillouin zone was sampled using a 19x19x11 **k**-point grid. The exchange-correlation was obtained using the Perdew-Burke-Ernzerhof parametrization of GGA [54]. We considered two structures in our calculations, $CrTe_2$ ($\delta$=0) and CrTe ($\delta$=1) – having $P6_3/mmc$ and $P3m1$ space group symmetries respectively. The atomic structures and lattice constants were relaxed using collinear magnetic configurations with a force component cutoff of 0.01 eV/Å. The magnetic exchange interactions were then calculated using the frozen magnon approach, with spin spirals generated on a 15x15x9 **q**-point grid using the generalized Bloch theorem and a spin spiral angle of $\pi/4$. The total energies obtained from these calculations give the exchange coupling in **q**-space J(**q**) and these can be Fourier transformed to generate exchange couplings in real space $J(\boldsymbol{R})$.

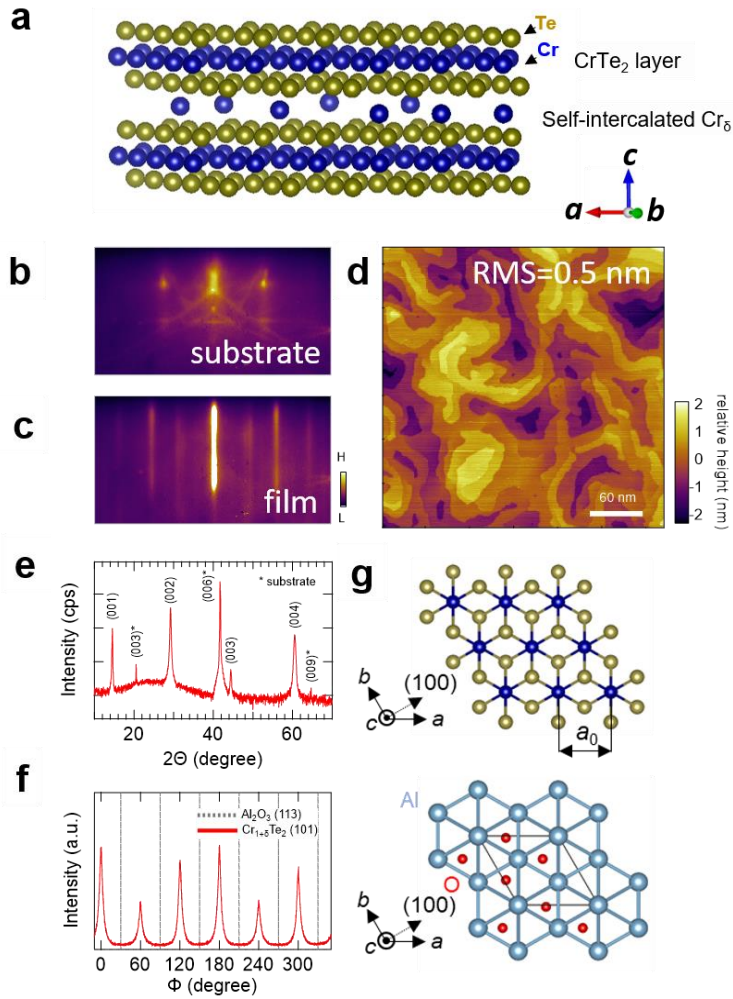

FIG. 1. Epitaxial growth of as–grown $Cr_{1+\delta}Te_2$ Film. (a) Side view of crystal structure of $Cr_{1+\delta}Te_2$ – showing intercalation of $Cr_\delta$ between $CrTe_2$ layers forming NiAs structure. (c) RHEED patterns for $Al_2O_3$ substrate – taken just before film deposition at 300°C and for as-grown film, taken at room temperature. The electron beam is injected along (100) direction of $Al_2O_3$ substrate. (d) STM topography (set bias: 200 mV, feedback current: 200 pA), with the root mean square (RMS) value of surface roughness indicated. (e) XRD profile along (00L) orientation of the as-grown film. (f) XRD φ scans with respect to asymmetric peaks – (101) of $Cr_{1+\delta}Te_2$ (red) and (113) of $Al_2O_3$ (gray). (g) Relative epitaxial orientation between substrate and film in this study.

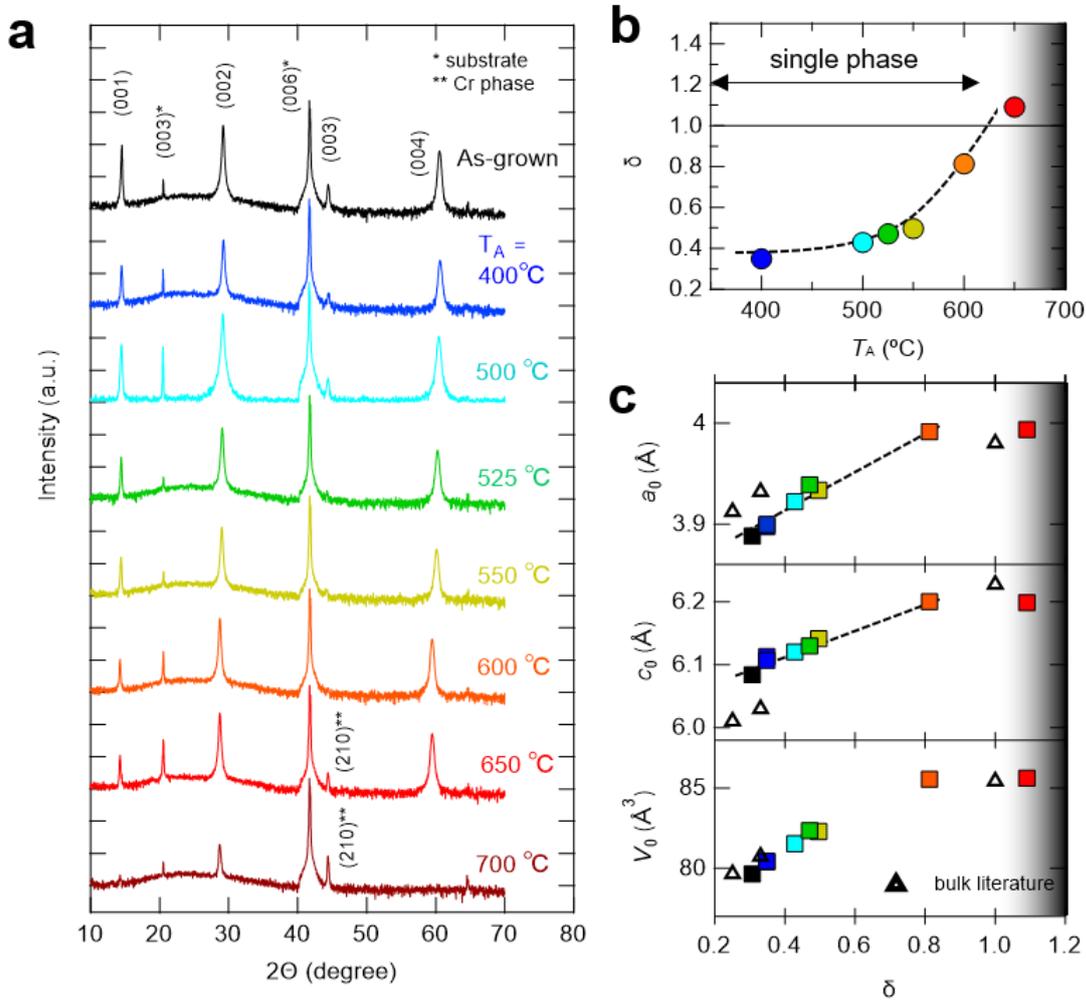

FIG. 2. Structural and compositional characterization after in-situ post deposition annealing with varying temperature ($T_A$). (a) XRD intensity profiles along (00$L$) direction for varying TA – vertically offset for clarity. * and ** indicate diffraction peaks originating from substrate and pure Cr, respectively. (b) SEM-EDS measurements of variation of δ as a function of $T_A$. The dashed parabolic trendline is a guide-to-the-eye. (c) variation of lattice constants $a_0$ and $c_0$ and unit cell volume $V_0$ as a function of δ – determined from XRD measurements. Dashed lines show a fit to Vegard's law. Triangular symbols indicate corresponding parameters for bulk crystals from literature [14,16,19].

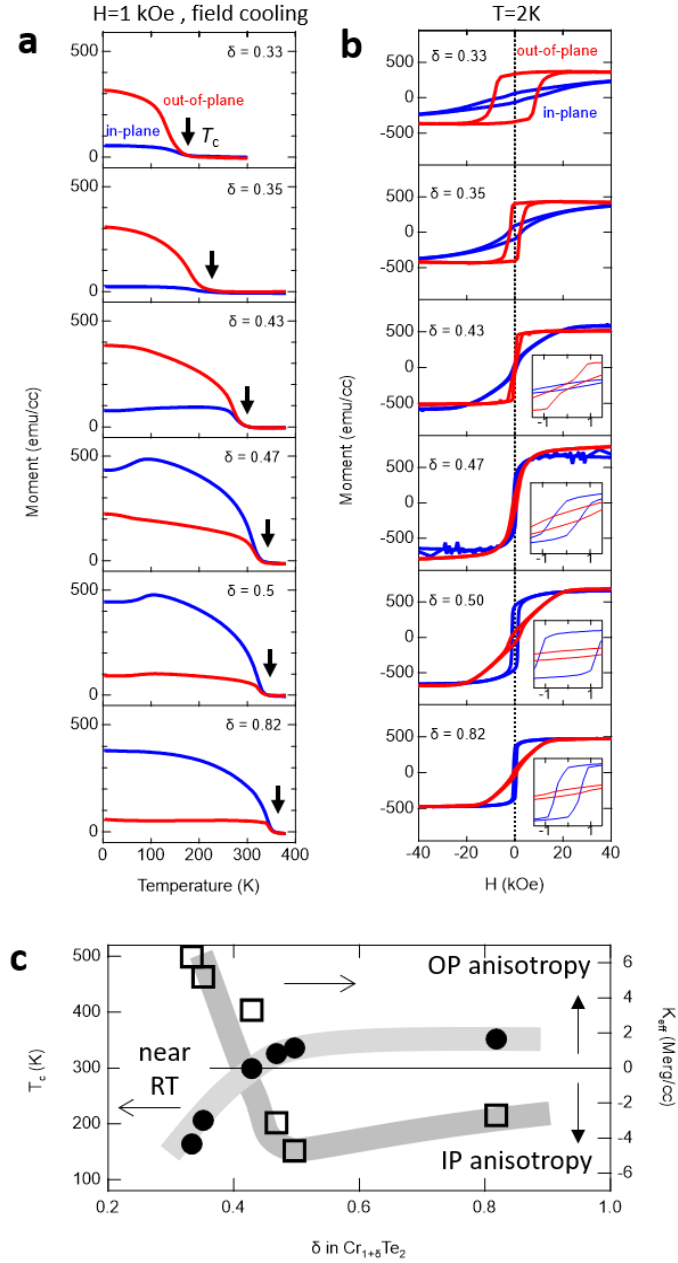

FIG. 3. Evolution of magnetic properties for $Cr_{1+\delta}Te_2$ films ($0.3 < \delta < 0.82$). Magnetization as a function of temperature, acquired by field cooling (FC) with a 1000 Oe magnetic field ($H$). Red and blue lines indicate data taken with $H$ along in out-of-plane (OP) and in-plane (IP) configurations, respectively. The Curie temperature $T_C$ is indicated by an arrow. (b) Magnetization hysteresis curves at 2K for each $\delta$, acquired in OP (red) and IP (blue) configurations, respectively. The diamagnetic signal from substrate is subtracted in (b). The effective magnetic anisotropy $K_{eff}$ was calculated from areal difference of IP and OP M(H) loops [45]. (c) The $\delta$ dependence of $T_C$ (left axis, filled circle) and $K_{eff}$ (right axis, empty square). Thick lines in (c) represent guides to the eye for $T_c$ and $K_{eff}$.

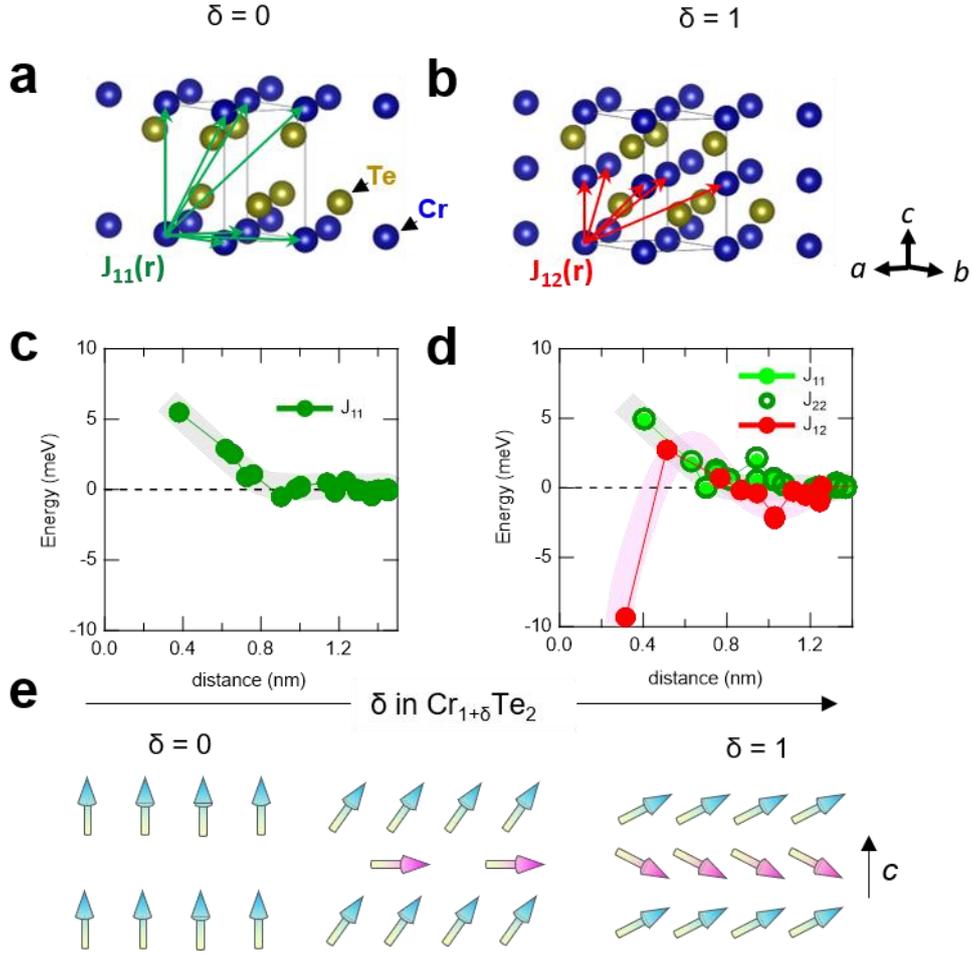

FIG. 4. Calculated exchange interactions for $Cr_{1+\delta}Te_2$ ($\delta$=0,1). (a-b) Crystal structures of $CrTe_2$ ($\delta$=0) and CrTe ($\delta$=1). Green and red arrows indicate direct exchange interactions for pairs of atoms in Intra-**s**ublattice ($J_{11}$) and inter-sublattice ($J_{12}$) configurations. For simplicity, (b) does not show green arrows for the two distinct intra-sublattice contributions, $J_{11}$ and $J_{22}$ that are included in calculation for CrTe (see d). **(c-d)** DFT-calculated exchange interactions, $J_{11}$, $J_{22}$ and $J_{12}$ as a function of atomic separation– for $\delta$=0 (c) and 1 (d) as per the structures in (a) and (b) respectively. Thick lines in (c) and (d) represent guides to the eye for distance evolution for $J_s$. **(e)** Cartoon showing expected evolution of ground state magnetic configuration of $Cr_{1+\delta}Te_2$ with varying $\delta$, with individual arrows indicating the local orientation of magnetization.

**Reference**

[1] A. Manchon, H. C. Koo, J. Nitta, S. M. Frolov, R. A. Duine, Nature Materials **14**, 871 (2015).

[2] F. Hellman *et. al.* Rev. Mod. Phys. **89**, 025006 (2017).

[3] A. Soumyanarayanan, N. Reyren, A. Fert, C. Panagopoulos, Nature **539**, 509 (2016).

[4] D. Akinwande, C. Huyghebaert, C. H. Wang, M. I. Serna, S. Goossens, L. J. Li, H.-S. Wong, F. H. L. Koppens, Nature **573**, 507 (2019).

[5] J. K. Slauhter, Annual Review of Materials Research **39**, 277, (2019).

[6] D. Apalkov, B. Dieny, J. M. Slaughter, Proceedings of the IEEE **104**, 1797 (2016).

[7] N. Nagaosa, Y. Tokura, Nature Nanotechnology **8**, 899 (2013).

[8] J. Matsuno, N. Ogawa, K. Yasuda, F. Kagawa, W. Koshibae, N. Nagaosa, Y. Tokura, M. Kawasaki, Sci. Adv. **2**, e1600304 (2016).

[9] A. Soumyanarayanan, M. Raju, A. L. G. Oyarce, A. K. C. Tan, M-Y. Im, A. P. Petrovic, P. Ho, K. H. Khoo, M. Tran, M, C. K. Gan, F. Ernult, C. Panagopoulos, Nature Materials **16**, 898 (2017).

[10] Y. Ohuchi, J. Matsuno, N. Ogawa, Y. Kozuka, M. Uchida, Y. Tokura, M. Kawasaki, Nature commun. **9**, 213 (2018).

[11] M. Nakamura, D. Morikawa, X. Yu, F. Kagawa, T. Arima, Y. Tokura, M. Kawasaki, J. Phys. Soc. Jpn. **87**, 074704 (2018).

[12] Z. Zhang, P. Yang, M. Hong, S. Jiang, G. Zhao, J. Shi, Q. Xie, Y. Zhang, Nanotechnology **30**, 182002 (2019).

[13] C. Gong, X. Zhang, Science **363**, 706 (2019).

[14] G. I. Makovetskii, A. I. Galyas, G. M. Severin, K. I. Yanushkevich, Inorganic materials **32**, 846 (1996).

[15] A. Ohsawa, Y. Yamaguchi, N. Kazama, H. Yamaguchi, H. Watanabe, J. Phys. Soc. Jpn. **33**, 1303 (1972).

[16] T. Hirone, S. Chiba, J. Phys. Soc. Jpn. **15**, 1991 (1960).

[17] M. Yamaguchi, T. Hashimoto, J. Phys. Soc. Jpn. **32**, 635 (1972).

[18] T. Hashimoto, K. Hoya, M. Yamaguchi, I. Ichitsubo, J. Phys. Soc. Jpn. **31**, 679 (1971).

[19] H. Ipser, L. K. Komarek, K. O. Klepp, J. Less-Common Metals **92**, 265 (1983)

[20] K. Lukoschus, S. Kraschinski, C. Nather, W. Bensch, R. K. Kremer, J. Solid State Chem. **177**, 951 (2004).

[21] Y. Wang, J. Yan, J. Li, S. Wang, M. Song, J. Song, Z. Li, K. Chen, Y. Qin, L. Ling, H. Du, L. Cao, X. Luo, Y. Xiong, Y. Sun, Phys. Rev. B **100**, 024434 (2019).

[22] X. H. Luo, W. J. Ren, Z. D. Zhang, J. Magn. Magn. Mat. **445**, 37 (2018).

[23] D. C. Freitas, R. Weht, A. Sulpice, G. Remenyi, P. Strobel, R. Gay, J. Marcus, M. N. Regueiro, J. Phys.: Cond. Matt. **27**, 176002 (2015).

[24] J. Yan, X. Luo, G. Lin, F. Chen, J. Gao, Y. Sun, L. Hu, P. Tong, W. Song, Z. Sheng, W. Lu, X. Zhu, Y. Sun, Euro. Phys. Lett. **124**, 67005 (2018).

[25] X. Sun, W. Li, X. Wang, Q. Sui, T. Zhang, Z. Wang, L. Liu, D. Li, S. Feng, S. Zhong, H. Wang, V. Bouchiat, M. N. Regueiro, N. Rougemaille, J. Coraux, Z. Wang, B. Dong, X. Wu, T. Yang, G. Yu, B. Wang,Z. V. Han, X. Han, Z. Zhang, arXiv: 1909.09797 (2019).

[26] Y. Wen, Z. Liu, Y. Zhang, C. Xia, B. Zhai, X. Zhang, G. Zhai, C. Shen, P. He, R. Cheng, L. Yin, Y. Yao, M. Sendeku, Z. Wang, X. Ye, C. Liu, C. Jiang, C. Shan, Y. Long, J. He, Nano letters **20**, 3130 (2020).

27 T. Hamasaki, T. Hashimoto, Y. Yamaguchi, H. Watanabe, Solid State Communications **16**, 895 (1975).

28 A. F. Andersen, Acta Chemica Scandinavia **24**, 3495 (1970).

29 S. Polesya, S. Mankovsky, D. Benea, H. Ebert, W. Bensch, J. Phys. Cond. Matt. **22**, 156002 (2010).

30 G. Cao, Q. Zhang, M. Frontzek, W. Xie, D. Gong, G. E. Sterbinsky, R. Jin, PHYSICAL REVIEW MATERIALS **3**, 125001 (2019).

31 A. Roy, S. Guchhait, R. Dey, T. Pramanik, C. C. Hsieh, A. Rai, S. K. Banerjee, ACS Nano **9**, 3772 (2015).

32 L. Zhou, J. S. Chen, Z. Z. Du, X. S. He, B. C. Ye, G. P. Guo, H. Z. Lu, G. Wang, H. T. He, *AIP Advances* **7**, 125116 (2017).

33 T. Pramanik, A. Roy, R. Dey, A. Rai, S. Guchhait, H. C. P. Movva, C. C. Hsieh, S. K. Banerjee, J. magne. Magne. Mat. **437**, 72 (2017).

34 D. M. Burn, L. B. Duffy, R. Fujita, S. L. Zhang, A. I. Figueroa, J. H. Martin, G. van der Laan, T. Hesjedal, Sci. Rep. **9**, 10793 (2019).

35 H. Li, L. Wang, T. Yu, L. Zhou, Y. Qiu, H. He, F. Ye, I. K. Sou, G. Wang, ACS Appl. Nano Mater. **2**, 6809 (2019).

36 D. Zhao, L. Zhang, I. A. Malik, K. Liao, W. Cui, X. Cai, C. Zheng, L. Li, X. Hu, D. Zhang, X. Chen, X. Jiang, Q. Xue, Nano Research **11**, 3116 (2018).

37 J. Chen, L. Wang, M. Zhang, L. Zhou, R. Zhang, L. Jin, X. Wang, H. Qin, Y. Qiu, J. Mei, F. Ye, B. Xi, H. He, B. Li, G. Wang, Nano Letters **19**, 6144 (2019).

38 L. Zhou, J. Chen, X. Chen, B. Xi, Y. Qiu, J. Zhang, L. Wang, R. Zhang, B. Ye, P. Chen, X. Zhang, G. Guo, D. Yu, J. W. Mei, F. Ye, G. Wang, H. He, *arXiv:* 1903.06486v1 (2019).

39 Y. Okada, Y. Ando, R. Shimizu, E. Minamitani, S. Shiraki, S. Watanabe, T. Hitosugi, Nature Commun. **8**, 15975 (2017).

40 M. Nakano, Y. Wang, S. Yoshida, H. Matsuoka, Y. Majima, K. Ikeda, Y. Hirata, Y. Takeda, H. Wadati, Y. Kohama, Y. Ohigashi, M. Sakano, K. Ishizaka, Y. Iwasa, Nano Letters **19**, 8806 (2019).

41 M. Nakano, Y. Wang, Y. Kashiwabara, H. Matsuoka, Y. Iwasa, Nano Letters **17**, 5595 (2017).

42 Y. Umemoto, K. Sugawara, Y. Nakata, T. Takahashi, T. Sato, Nano Research **12**, 165 (2019).

43 R. Watanabe, R. Yoshimi. M. Shirai, T. Tanigaki, M. Tsukazaki, K. S. Takahashi, R. Arita, M. Kawasaki, Y. Tokura, Appl. Phys. Lett. **113**, 181602 (2018).

44 A. R. Denton, N. W. Ashcroft, Phys. Rev. A **43**, 3161 (1991).

45 M. T. Johnson, P. J. H. Bloemen, F. J. A. den Broeder, J. J. de Vries, Rep. Prog. Phys. **59**, 1409 (1996).

46 N. Majlis, The Quantum Theory of Magnetism (Singapore: World Scientific) chapter 3

47 B. Dupé, M. Hoffmann, C. Paillard, S. Heinze, Nat. Commun **5**, 4030 (2014).

48 T. Kurumaji, T. Nakajima, M. Hirschberger, A. Kikkawa, Y. Yamasaki, H. Sagayama, H. Nakao, Y. Taguchi, T. H. Arima, Y. Tokura, Science **365**, 914 (2019).

49 K. Monna, F. Izumij, J. Appl. Crystallogr. **44**, 1272 (2011).

50 R. Laskowski, G. K. H. Madsen, P. Blaha, K. Schwarz, Phys. Rev. B **69**, 140408(R) (2004).

51 J. Kunes, R. Laskowski, Phys. Rev. B **70**, 174415 (2004).

52 P. Blaha, K. Schwarz, F. Tran, R. Laskowski, G. K. H. Madsen, L. D. Marks, J. Chem. Phys. **152**, 074101 (2020).

53 J. P. Perdew, K. Burke, M. Ernzerhof, Phys. Rev. Lett. **77**, 3865 (1996).

54 L. M. Sandratskii, Adv. Phys. **47**, 91 (1998).